\documentclass{emulateapj}
\usepackage{amsmath}
\usepackage{epstopdf}
\usepackage{apjfonts}
\usepackage{xspace}
\usepackage{hyperref} 
\usepackage{color}
\hypersetup{%
    colorlinks=false, 
    urlcolor=magenta, 
    linkcolor=blue, 
    citecolor=blue,
    breaklinks=true, 
    bookmarksnumbered=true,
    bookmarksopen=false, 
    bookmarksopenlevel=0,
    hypertexnames=true, 
    unicode=true,          
    pdftoolbar=true,        
    pdfmenubar=true,        
    pdffitwindow=false,     
    pdfstartview={FitH}    
}

\usepackage{etoolbox}
\makeatletter

\patchcmd{\NAT@citex}
  {\@citea\NAT@hyper@{\NAT@nmfmt{\NAT@nm}\NAT@date}}
  {\@citea\NAT@nmfmt{\NAT@nm}\NAT@hyper@{\NAT@date}}
  {}
  {}

\patchcmd{\NAT@citex}
  {\@citea\NAT@hyper@{%
     \NAT@nmfmt{\NAT@nm}%
     \hyper@natlinkbreak{\NAT@aysep\NAT@spacechar}{\@citeb\@extra@b@citeb}%
     \NAT@date}}
  {\@citea\NAT@nmfmt{\NAT@nm}%
   \NAT@aysep\NAT@spacechar%
   \NAT@hyper@{\NAT@date}}
  {}
  {}

\patchcmd{\NAT@citex}
  {\@citea\NAT@hyper@{%
     \NAT@nmfmt{\NAT@nm}%
     \hyper@natlinkbreak{\NAT@spacechar\NAT@@open\if*#1*\else#1\NAT@spacechar\fi}%
       {\@citeb\@extra@b@citeb}%
     \NAT@date}}
  {\@citea\NAT@nmfmt{\NAT@nm}%
   \NAT@spacechar\NAT@@open\if*#1*\else#1\NAT@spacechar\fi%
   \NAT@hyper@{\NAT@date}}
  {}
  {}

\makeatother

\newcommand{\ysz}{$Y$}
\newcommand{\yszth}{$Y_{th}$}
\newcommand{\ysznt}{$Y_{nt}$}
\newcommand{\ysztot}{$Y_{th} + Y_{nt}$}

\begin{document}
\submitted{Accepted to ApJ May 21, 2015}
\slugcomment{Accepted to ApJ May 21, 2015} 

\title{The Influence of Mergers on Scatter and Evolution in Sunyaev-Zel'dovich Effect Scaling Relations}
\author{Liang Yu\altaffilmark{1,2}}
\author{Kaylea Nelson\altaffilmark{1,3}}
\author{Daisuke Nagai\altaffilmark{1,2,3}}

\affil{ 
$^1$Department of Astronomy, Yale University, New Haven, CT 06520, U.S.A. \\ 
$^2$Department of Physics, Yale University, New Haven, CT 06520, U.S.A.\\
$^3$Yale Center for Astronomy \& Astrophysics, Yale University, New Haven, CT 06520, U.S.A. \\
}


\begin{abstract}
The Sunyaev-Zel'dovich effect (SZE) observable-mass ($Y-M$) scaling relation is a promising technique for obtaining mass estimates for large samples of galaxy clusters and holds a key to studying the nature of dark matter and dark energy. However, cosmological inference based on SZE cluster surveys is limited by our incomplete knowledge of scatter, and evolution in the $Y-M$ relation. In this work, we investigate the effects of galaxy cluster mergers on the scaling relation using the {\it Omega500} high-resolution cosmological hydrodynamic simulation. We show that the non-thermal pressure associated with merger-induced gas motions contributes significantly to the scatter, and evolution of the scaling relation. After the merger, the kinetic energy of merging systems is slowly converted into thermal energy through dissipation of turbulent gas motions, which causes the thermal SZE signal to increase steadily with time. This post-merger evolution is one of the primary source of scatter in the $Y-M$ relation. However, we show that when the missing non-thermal energy is accounted for, the resulting relation exhibits little to no redshift evolution and the scatter around the scaling relation is $\sim 20-30 \%$ smaller than that of the thermal SZE signal alone. Our work opens up a possibility to further improve the current robust mass proxy, $Y$, by accounting for the missing non-thermal pressure component. We discuss future prospect of measuring internal gas motions in galaxy clusters and its implication for cluster-based cosmological tests.
\end{abstract}

\section{Introduction}
\label{Introduction}

Galaxy clusters are the largest gravitationally collapsed objects in the universe. Their formation and evolution are critically sensitive to details of the expansion history of the universe. Accurate measurements of their abundance as a function of cluster mass and redshift can therefore provide important constraints on cosmological parameters such as $\Omega_M$, $\Omega_{\Lambda}$, $\sigma_8$ and $w_0$ (see, e.g. \citealt{vikhlinin09,mantz10}).  However, the use of galaxy clusters as cosmological probes is currently limited by our ability to relate observable properties and cluster mass through so-called observable-mass relations.

In recent years, the Sunyaev-Zel'dovich effect (SZE) has emerged as a promising avenue for both detecting clusters and obtaining robust mass estimates for hundreds of clusters in the high-redshift universe \citep[][for review]{carlstrom02}. The SZE is a small distortion in the cosmic microwave background (CMB) spectrum caused by the scattering of CMB photons off a distribution of high-energy electrons in dense structures such as clusters of galaxies \citep[][for review]{sz70, sz72, birkinshaw99}. The SZE has the unique property that its signal is independent of redshift and does not suffer from cosmological surface brightness dimming, in contrast to X-ray observations which have been the primary observational tools for studying the hot X-ray emitting intracluster medium (ICM) in galaxy clusters.  Moreover, the integrated SZE signal, $Y$, is only weakly dependent on nonlinear astrophysical processes (e.g. cooling flows and energy feedback from supernova and supermassive black holes) which play an important role in determining the properties of cluster cores and significantly affect their X-ray luminosity, making the SZE an excellent probe of cluster mass. 

The current generation of SZE cluster surveys, conducted with the Atacama Cosmology Telescope (ACT), the South Pole Telescope (SPT), and the Planck space mission, have obtained large ($\sim 10^4$) and statistically representative samples of massive clusters across the full sky \citep{hasselfield13,reichardt13,Planck2014XXIX} out to a median redshift significantly higher than that of X-ray selected cluster catalogs.  However, to realize the full statistical power of SZE surveys, systematic uncertainties in the $Y-M$ relation must be controlled at a level comparable to statistical uncertainties. For example, Planck Collaboration has recently reported a tension between the cosmological constraints from CMB and galaxy clusters \citep{Planck2014XX}, possibly indicating an unexpectedly large bias in the calibration of the $Y-M$ relation \citep[but see also][]{rozo14}. A detailed study of the $Y-M$ relation is, therefore, particularly timely and important. 

Theoretically, modern hydrodynamical simulations suggest that a SZE signal integrated to a sufficiently large fraction of cluster volume is a robust mass proxy with a scatter of $10-15 \%$ \citep{motl05, nagai06, stanek10, Sembolini2012}. Specifically, while the normalization is sensitive to cluster astrophysics such as gas cooling, star formation, and energy feedback from stars and active galactic nuclei (AGN), the scatter and evolution of the $Y-M$ relation is fairly insensitive to the poorly understood cluster astrophysics \citep{nagai06, battaglia11, fabjan11}.  Cluster merger dynamics, on the other hand, has been shown to play an important role in determining the scatter and evolution of the $Y-M$ relation \citep{wik08, yang10,krause12}.  

This picture appears to be supported by recent observations which showed that dynamically disturbed clusters have increased scatter around the relation compared to their more relaxed counterparts \citep{sifon13}. To harness the potential of the SZE surveys, we must understand the origin of scatter, and evolution in the $Y-M$ relation and devise a technique to control their systematic uncertainties to better than a few percent for a wide range of masses and redshifts.

In this work, we use the {\it Omega500} high-resolution cosmological simulation to demonstrate that the scatter and evolution in the $Y-M$ relation are governed primarily by the evolution of the non-thermal pressure provided by internal gas motions generated by major mergers. To characterize the effects of mergers and dynamical states of clusters, we (1) analyze non-radiative simulations in which the ICM is heated by dominant gravitational heating processes (such as shock and compressional heating) alone and (2) use detailed merger trees to track each cluster's formation history and directly examine the effects of major mergers on the scaling relation. We first show that large deviations from the scaling relation in unrelaxed systems are due to a large influx of non-thermal pressure from random gas motions generated by major mergers. We also find that the scaling relation departs from self-similarity at high redshift where mergers are more ubiquitous.  By accounting for the missing non-thermal energy contribution, we demonstrate explicitly that the thermal plus non-thermal pressure exhibits remarkable regularity and reduced scatter at all redshifts, indicating that major mergers are a dominant source of uncertainty in this relation.

This paper is organized as follows: in Section~\ref{sec:theory} we describe the thermal SZE and present relevant scaling relations predicted by the self-similar model, as well as previous results on the evolution of thermal pressure support after major mergers; in Section~\ref{sec:simul} we describe the {\it Omega500} cosmological simulation; in Section~\ref{sec:results} we present our analysis and interpretation of the evolution of SZE-mass scaling relations; finally, in Section~\ref{sec:conc}, we summarize our findings and discuss their implications for current and future SZE cluster surveys.

\section{Theoretical framework}
\label{sec:theory}

\subsection{Thermal SZE}

The thermal SZE is a distortion in the CMB spectrum produced by the
inverse Compton scattering of CMB photons off free electrons in dense
structures such as clusters of galaxies \citep{sz70, sz72}. This scattering causes a small change in the mean photon energy as $\Delta \nu/\nu \simeq k_BT/(m_ec^2) \sim 10^{-2}$, where $\nu$ is the frequency of the CMB photon and $m_e$ is the electron rest mass. As a result, the intensity of the CMB spectrum increases in the high-frequency (Wien) end and decreases in the low-frequency (Rayleigh-Jeans) tail. The corresponding brightness fluctuation in the CMB is of order $\sim 10^{-4}$,  about an order of magnitude larger than the fluctuations from primary anisotropies.

The change in the CMB specific intensity at a frequency $\nu$ caused by the thermal SZE is proportional to the line-of-sight integral of the number density ($n_e$) and temperature ($T_e$) of electrons, and is given by $\Delta I_{\nu}/I_{\rm CMB} = f_{\nu}(x)g_{\nu}(x)y$, where $x \equiv h\nu/k_BT_{\rm CMB}$, $f_{\nu}(x)= x(e^x+1)/(e^x-1)-4$, and $g_{\nu}(x) = x^4 e^x/(e^x-1)^2$. The dimensionless Comptonization parameter $y$ is defined as
\begin{equation}
y \equiv \frac{k_B \sigma_T}{m_e c^2} \int n_e(l) T_e(l) dl
\end{equation}
where $c$ is the speed of light, and $\sigma_T$ is the Thomson cross-section.  This equation corresponds to the integrated line-of-sight thermal pressure of the intracluster gas, which is approximated as $P=n_ek_BT$ in the ideal gas limit. The corresponding change in the CMB temperature is given by $\Delta T_{\nu}/T_{\rm CMB}=f_{\nu}(x) y$.  In
the Rayleigh-Jeans limit ($\nu \ll 200$GHz), $\Delta T_{\nu}/T_{\rm
CMB}= -2y$ and $\Delta I_{\nu}=(2k_B \nu^2/c^2) \Delta T_{\nu}$.

Let us now consider the SZE signal arising from a cluster located at
redshift $z$.  The SZE flux integrated across the surface of a cluster is defined as the integrated Compton-y parameter $Y_{SZ}$:
\begin{equation}
\label{eq:Y}
Y \equiv \int_{\Omega} y d\Omega = \frac{1}{d_A^2(z)} \left(\frac{k_B
\sigma_T}{m_e c^2}\right) \int_V n_e(l) T_e(l) dV,
\end{equation}
and $d\Omega = dA/d_A^2(z)$ is the solid angle of the cluster
subtended on the sky (the integral is taken over the volume of the cluster), $d_A(z)$ is the angular diameter distance to the cluster, $dA$ is the area of the cluster on the sky, and $dV$ is the cluster volume. $Y$ measures the total thermal energy of a cluster.

The thermal SZE signal is linearly sensitive to gas mass $M_{\rm gas}=f_{gas}M$ and mass-weighted temperature $T_m$:
\begin{equation}
\label{eq:Yint}
Y \propto f_{\rm gas} M T_m.
\end{equation}
where $f_{gas}$ is the gas mass fraction and $M$ is the total cluster mass. In this work, we use a spherically integrated $Y$ within a radius of interest around the cluster center, rather than the projected $Y$ derived from observations. This enables our work to focus on the direct effects of mergers and dynamical state on $Y$.

\subsection{Self-similar Scaling Relations}
In the absence of cooling and heating processes, clusters are expected to scale self-similarly \citep{kaiser86}.  The self-similar model predicts that the temperature of the gas scales with the cluster mass as
\begin{equation}
\label{eq:mt} 
M \propto T_m^{3/2} E^{-1}(z)
\end{equation} 
where $M \equiv 4\pi r^3_{\Delta} \Delta_{c} \rho_{crit}/3$ is the halo mass enclosed within $r_{\Delta}$, defined as a radius of spherical volume within which the mean density is $\Delta_{c}$ times the {\it critical density}, $\rho_{\rm crit}$, at that redshift \citep{bryan98}.  $E(z)$ is the redshift-dependent Hubble parameter, defined as $H(z)=100hE(z)$ km s$^{-1} {\rm Mpc}^{-1}$, and it is given by $E^2(z)=\Omega_M(1+z)^3+\Omega_{\Lambda}$ for a flat cosmology.

Inserting Eq.~\ref{eq:mt} into Eq.~\ref{eq:Yint}, we obtain the SZE
scaling relation predicted by the self-similar model,
\begin{equation}
\label{eq:Yscale}
Y \propto f_{\rm gas} M^{5/3} E^{2/3}(z),
\end{equation}
where $f_{gas}$ is taken to be constant in the self-similar model, such that $Y \propto M^{5/3} E^{2/3}(z)$.  In practice, we fit a straight line to the $\ln Y - \ln M$ relation by minimizing $\chi ^2$, where the best-fit relation is given by
\begin{equation}
\label{eq:ymod}
\ln Y_{fit} = \ln A_{14} + \alpha \ln \left(\frac{M}{10^{14} h^{-1}M_{\odot}}\right)  + \beta \ln E(z),
\end{equation}
where $\ln A_{14}$ is the normalization constant of the best-fit relation at $M=10^{14} h^{-1}M_{\odot}$, and $\alpha$ is the best-fit slope of the $Y-M$ relation. Note that the self-similar model predicts $\alpha=5/3$. $\beta$ is the redshift dependence of the relation and is fixed to the self-similar value of 2/3.

Throughout this paper, we consider $Y$, mass and other cluster properties within two commonly used radii, defined by the total matter overdensity they enclose and chosen for their relevance to X-ray and SZ observations. We use radii $r_{500c}$ and $r_{200c}$, enclosing overdensities of $\Delta_c$ = 500 and 200 times the critical density of the universe, $\rho_{\rm crit}$ and radius  $r_{200m}$ enclosing overdensities of $\Delta_m = 200$ times the mean density of the universe.

\subsection{Evolution of Thermal and Non-thermal Pressure Content}
\label{subsec:e_of_nt}
The evolution of thermal and non-thermal pressure support after mergers has been studied extensively by \citet{nelson12}. \citet[hereafter N12]{nelson12} have shown that during a major merger, non-thermal pressure from random gas motions provide about a third of the total gas pressure in the ICM.  As clusters relax, the non-thermal pressure support falls off rapidly, ultimately comprising only a residual 10\% contribution. This decay of non-thermal pressure support is due to the conversion of random velocities to thermal energy via shocks and dissipation. As a result, the total energy content (thermal and non-thermal) of the ICM remains nearly constant after a merger -- a decrease in random kinetic energy corresponds to an increase in thermal energy \citep{vazza11, nelson12}. In this paper, we consider properties of the ICM measured under the assumption of spherical symmetry following the methods described in \citet{lau_etal09}. The total energy content of cluster gas has two components: the thermal energy from the hot ICM and non-thermal energy from gas motions, herein defined respectively as 

\begin{equation}
\varepsilon_{\rm th} = \int P_{th} \ dV
\end{equation}

\begin{equation}
\varepsilon_{\rm nt} = \int \frac{1}{3} \rho_{\rm gas}(\sigma^2_r + \sigma^2_t)\ dV
\end{equation}
where $P_{th}$ is the thermal pressure, $\rho_{\rm gas}$ is the gas density, and $\sigma^2_r$ and $\sigma^2_t$ are radial and tangential dispersions measured in spherical shells, using the peculiar velocity of the cluster dark matter within $r_{500c}$ to define a rest frame. However, we have verified that our results are insensitive to the choice of rest frame. Note that a complete description of non-thermal energy includes rotation, streaming and cross terms. However, these contributions are small compared to that of random motions \citep{lau13} and neglected in this work.

As stated above, the $Y-M$ scaling relation is dependent on a tight correlation between the thermal properties of the ICM and the mass of the cluster.  However, as shown in N12, a large amount of non-thermal pressure is injected into a cluster during a major merger which only thermalizes over a few Gyr timescale, resulting in a prolonged deviation from this relation. To characterize the contribution from the missing non-thermal energy contribution,  we calculate a combined $Y$,

\begin{equation}
Y_{th} + Y_{nt} = \frac{\varepsilon_{th} + \varepsilon_{nt}}{\varepsilon_{th}} Y_{th},
\label{eq:ss}
\end{equation}

\noindent where $Y_{th}$ and $Y_{nt}$ are the thermal and non-thermal components of $Y$. \ysztot\ is proportional to the total energy content of the ICM and therefore more closely correlates with the total mass of the system. Note that spherical symmetry for the gravitational potential and steady state are assumed in deriving the expression for non-thermal energy, and all the physical quantities at a given radius are averages over a radial shell \citep{lau13}.

\section{Simulations}
\label{sec:simul}
\subsection{Hydrodynamic Simulations of Galaxy Clusters}
\label{subsec:hsimul}

We analyze the {\it Omega500} simulation of galaxy clusters presented previously in \citet{nelson14b}. We refer the reader to that paper for more details. We briefly summarize the relevant parameters below.

In this work, we analyze the {\it Omega500} high-resolution cosmological simulation of massive galaxy clusters in a flat $\Lambda$CDM model with WMAP five-year ({\em WMAP5}) cosmological parameters: $\Omega_M = 1 - \Omega_{\Lambda} = 0.27$, $\Omega_b = 0.0469$, $h = 0.7$ and $\sigma_8 = 0.82$, where the Hubble constant is defined as $100h$~km~s$^{-1}$~Mpc$^{-1}$ and $\sigma_8$ is the mass variance within spheres of radius 8$h^{-1}$ Mpc. The simulation is performed using the Adaptive Refinement Tree (ART) $N$-body+gas-dynamics code \citep{kravtsov99, kravtsov02, rudd_etal08}, which is an Eulerian code that uses adaptive refinement in space and time, and non-adaptive refinement in mass \citep{klypin_etal01} to achieve the dynamic ranges to resolve the cores of halos formed in self-consistent cosmological simulations. The simulation volume has a comoving box length of $500\,h^{-1}$~Mpc, resolved using a uniform $512^3$ grid and 8 levels of mesh refinement, implying a maximum comoving spatial resolution of $3.8\,h^{-1}$~kpc.  

We selected clusters with $M_{500c} \geq 2.9\times10^{14} h^{-1}M_{\odot}$ and performed a simulation where only the regions surrounding the selected clusters are resolved. The resulting simulation has effective mass resolution of $2048^{3}$ surrounding the selected clusters, allowing a corresponding mass resolution of $1.09 \times 10^9\, h^{-1}M_{\odot}$. The current simulation only models gravitational physics and non-radiative hydrodynamics. \citet{nagai06} has studied the effect of cooling and star formation on the SZE-mass scaling relation, and concludes that cooling and star formation have a significant effect on the normalization but not on the scatter. Moreover, \citet{nelson14b} have also examined the effect of radiative cooling, star formation and energy feedback from SNe on the non-thermal pressure fraction and gas velocity anisotropy in group and cluster size halos and found little systematic dependence on gas physics.

To study the evolution of $Y$,  we extract halos from four redshift outputs: $z=0.0, 0.5,1.0, 1.5$. At each redshift we apply an additional mass-cut to ensure mass-limited samples at all epochs. The mass-cuts and resulting sample sizes are as follows: 65 clusters with $M_{200m} \geq 6\times 10^{14} h^{-1}M_{\odot}$ at $z = 0$,  48 clusters with $M_{200m} \geq 2.5\times 10^{14} h^{-1}M_{\odot}$ at $z = 0.5$,  42 clusters with $M_{200m} \geq 1.3\times 10^{14} h^{-1}M_{\odot}$ at $z = 1.0$,  and 42 clusters with $M_{200m} \geq 7 \times 10^{13} h^{-1}M_{\odot}$ at $z = 1.5$.

\label{subsec:tmerger}
\begin{figure*}[t]
\epsscale{1}
\plottwo{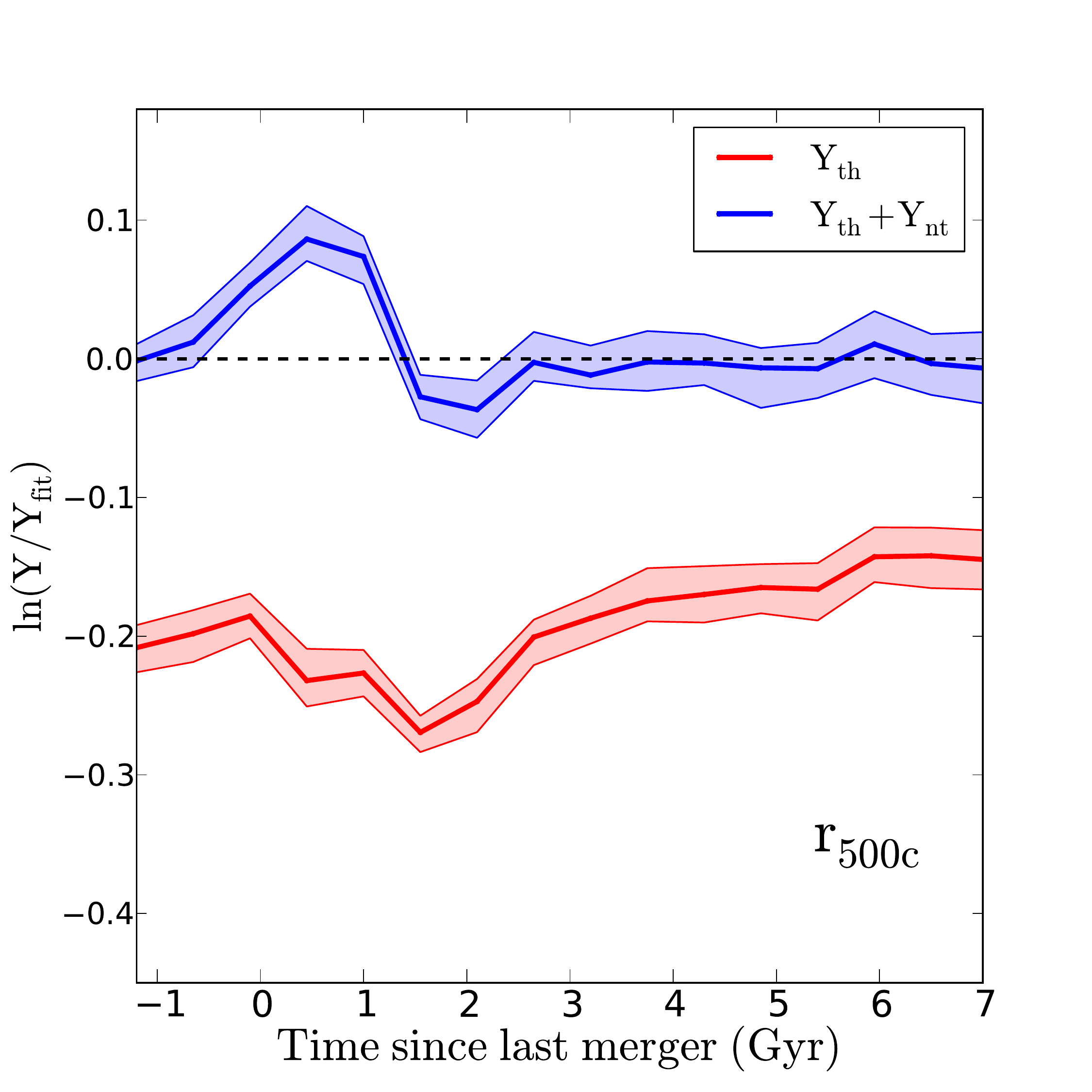}{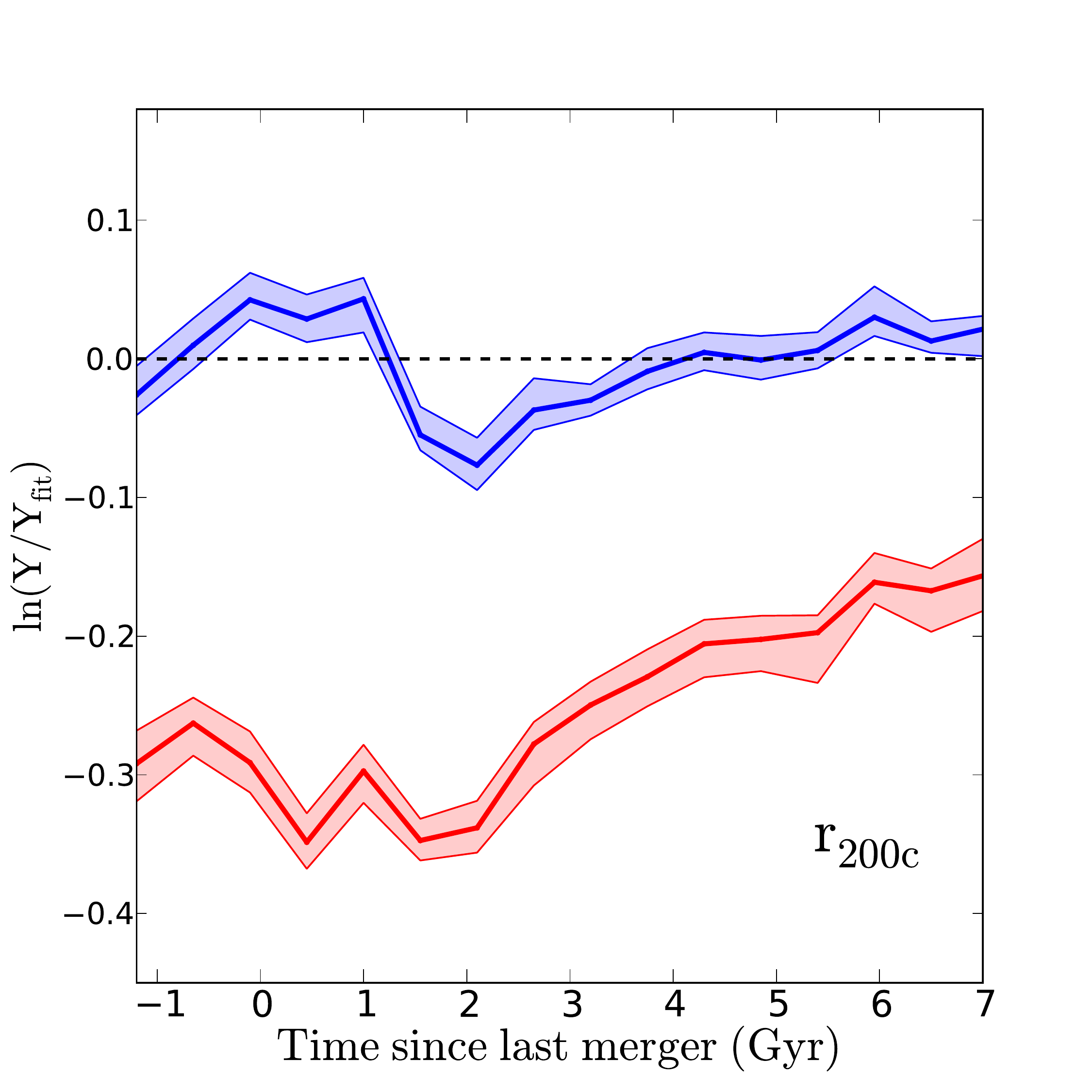}
  \caption{Mean fractional deviation of thermal \yszth\ ({\it red}) and \ysztot\ ({\it blue}) at $r_{500c}$ ({\it left}) and $r_{200c}$ ({\it right}) from the best-fit relation for \ysztot\, $Y_{fit}$, shown as a dashed horizontal line in each plot. The shaded regions represent $1\sigma$ deviation from the mean.}
\label{fig:4in1} 
\end{figure*}

\subsection{Merger Trees}

To examine the evolution of cluster properties, we track the evolution of each cluster's most massive progenitor. We identify the main progenitor of each $z = 0$ cluster by iteratively following the dark matter particles at each epoch.  Mergers are located by identifying objects that cross $r_{500c}$ of the parent cluster.  We determine the mass ratio of the merger at an epoch when the two systems are sufficiently separated to avoid impacting their respective mass profiles, herein we use the epoch at which they initially overlap at a radius of $r_{500c}$. We use the ratio of $M_{500c}$ at this epoch to identify major mergers, herein defined as a mass ratio greater than 1:6. We have also experimented with other merger ratios, but found that 1:6 gives the cleanest trends with the smallest cluster-to-cluster scatter. Merging events with mass ratios smaller than this are defined as minor mergers. While this ratio is somewhat lower than that used in other studies \citep[e.g.][]{gottlober01}, careful exploration of the merger ratio has shown that this threshold best separates mergers that have a strong impact on the measured ICM properties from those that do not. We examine the evolution of ICM properties as a function of time since the last major merger, defined as the epoch at which they initially overlap at a radius of $r_{500c}$.

\begin{deluxetable*}{lccccccc}
\tabletypesize{\scriptsize}
\setlength{\tabcolsep}{-2in} 
\tablewidth{0pt}
\tablecolumns{5}
\tablecaption{Summary of transit parameters }
\tablehead{
\multicolumn{1}{l}{} &
\multicolumn{3}{c}{$Y_{th}$} &
\multicolumn{1}{c}{} &
\multicolumn{3}{c}{$Y_{th}+Y_{nt}$} \\
\multicolumn{1}{l}{Redshift}&
\multicolumn{1}{c}{$\ln A_{14}$} &
\multicolumn{1}{c}{$\alpha$} &
\multicolumn{1}{c}{$\sigma$} &
\multicolumn{1}{c}{} &
\multicolumn{1}{c}{$\ln A_{14}$} &
\multicolumn{1}{c}{$\alpha$} &
\multicolumn{1}{c}{$\sigma$} 
}
\startdata
0 & $-12.01 \pm 0.07$ & $1.63 \pm 0.05$ & $0.114^{+0.023}_{-0.019}$ & & $-11.88 \pm 0.07$ & $1.68 \pm 0.05$ & $0.115^{+0.031}_{-0.023}$ \vspace{0.5mm} \\ 
0.5 & $-12.06 \pm 0.04$ & $1.64 \pm 0.04$ & $0.103^{+0.036}_{-0.023}$ & & $-11.87 \pm 0.03$ & $1.69 \pm 0.03$ & $0.090^{+0.034}_{-0.018}$ \vspace{0.5mm} \\
1 & $-12.12 \pm 0.03$ & $1.71 \pm 0.05$ & $0.127^{+0.033}_{-0.021}$ & & $-11.87 \pm 0.02$ & $1.76 \pm 0.04$ & $0.090^{+0.021}_{-0.017}$ \vspace{0.5mm} \\
1.5 & $-12.12 \pm 0.02$ & $1.60\pm 0.05$ & $0.113^{+0.032}_{-0.022}$ & & $-11.84 \pm 0.02$ & $1.64 \pm 0.03$ & $0.084^{+0.049}_{-0.022}$ \vspace{0.5mm} \\
0 (relaxed) & $ -12.06 \pm 0.06 $ & $1.71 \pm 0.04 $ & $0.070^{+0.007}_{-0.007}$ & & $-11.99 \pm 0.05$ & $1.68 \pm 0.05$ & $0.058^{+0.006}_{-0.005}$ \vspace{1.0mm} \\ 
\hline 
\vspace{-2.0mm} \\
0 &  $-12.11 \pm 0.10$ & $1.56 \pm 0.05$ & $0.123^{+0.023}_{-0.017}$ & & $-12.00 \pm 0.07$ & $1.62 \pm 0.04$ & $0.085^{+0.020}_{-0.015}$ \vspace{0.5mm} \\ 
0.5 & $-12.28 \pm 0.06$ & $1.58 \pm 0.05$ & $0.112^{+0.027}_{-0.020}$ & & $-12.06 \pm 0.04$ & $1.63 \pm 0.03$ & $0.082^{+0.017}_{-0.013}$ \vspace{0.5mm} \\
1 & $-12.43 \pm 0.04$ & $1.66 \pm 0.05$ & $0.121^{+0.026}_{-0.021}$ & & $-12.14\pm 0.03$ & $1.70 \pm 0.04$ & $0.085^{+0.025}_{-0.014}$ \vspace{0.5mm} \\
1.5 & $-12.42 \pm 0.02$ & $1.59\pm 0.05$ & $0.120^{+0.027}_{-0.022}$ & & $-12.09 \pm 0.02$ & $1.61 \pm 0.04$ & $0.100^{+0.018}_{-0.016}$ \vspace{0.5mm} \\
0 (relaxed) & $-12.12 \pm 0.10$ & $1.62 \pm 0.06$ & $0.085^{+0.010}_{-0.009}$ & & $-12.05 \pm 0.06$ & $1.67 \pm 0.03$ & $0.054^{+0.006}_{-0.006}$ \vspace{0.5mm}
\enddata
\normalsize
\label{tab:bestfit}
\end{deluxetable*}
\vspace{0.2cm}

\begin{figure*}[t]
\epsscale{0.90}
\plottwo{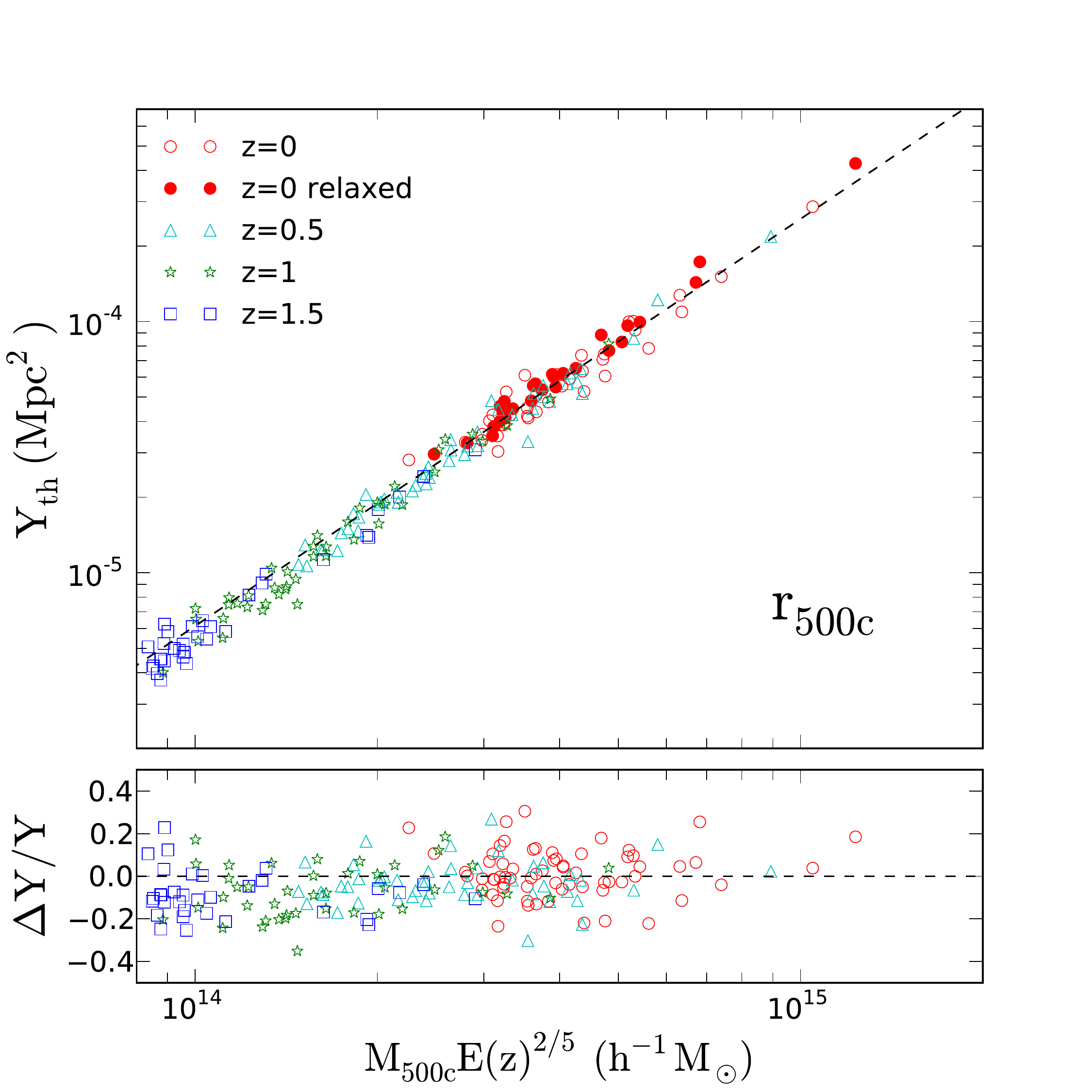}{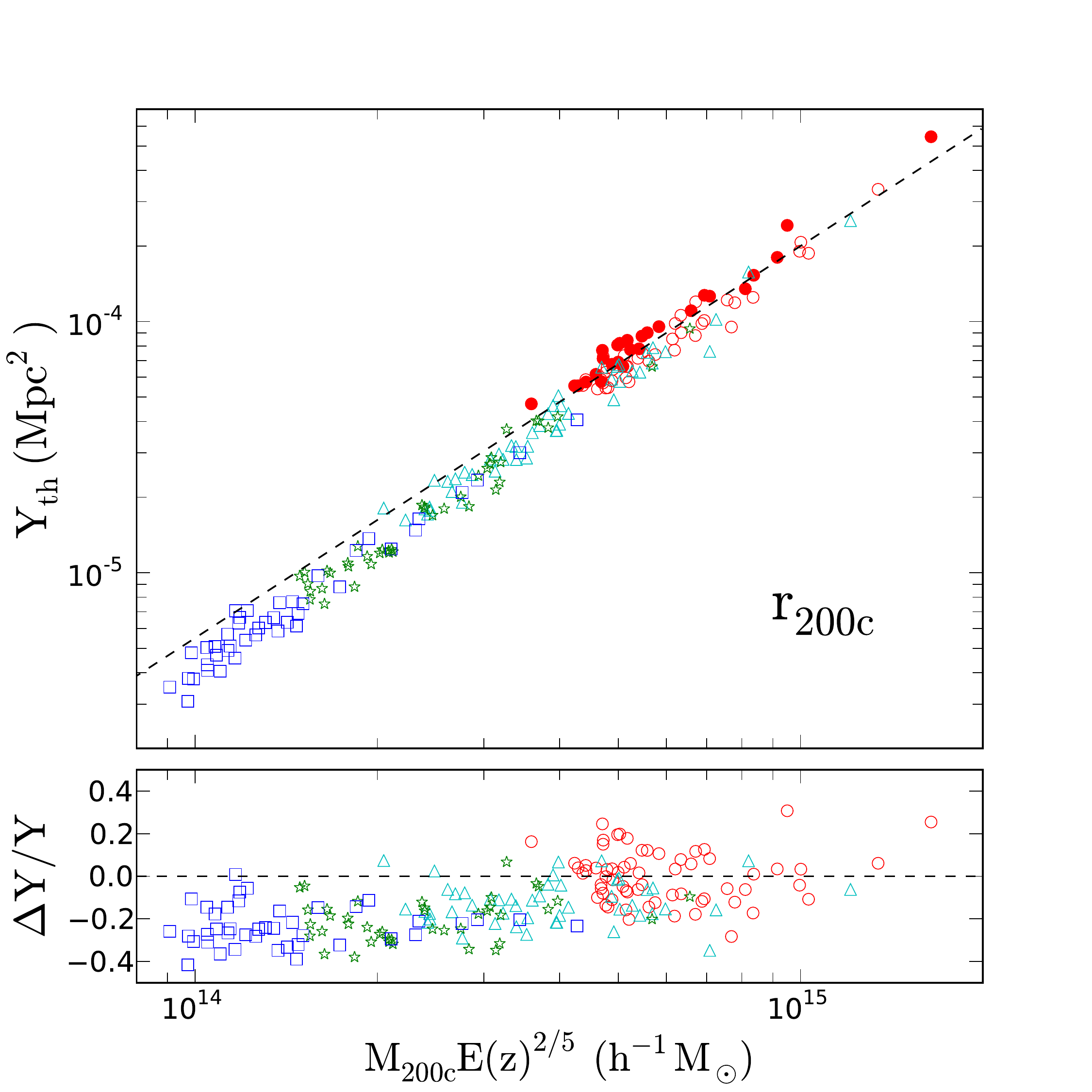}
\plottwo{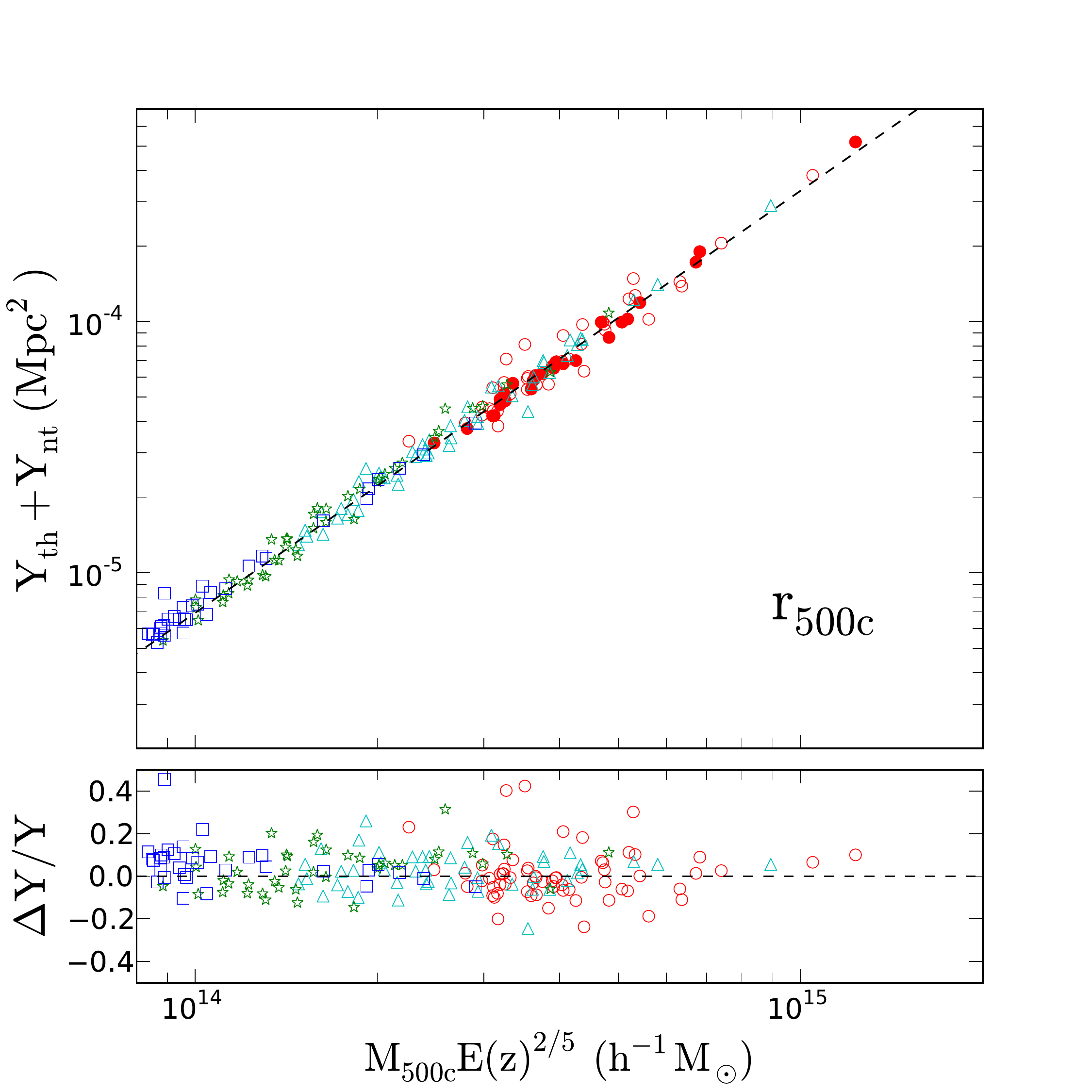}{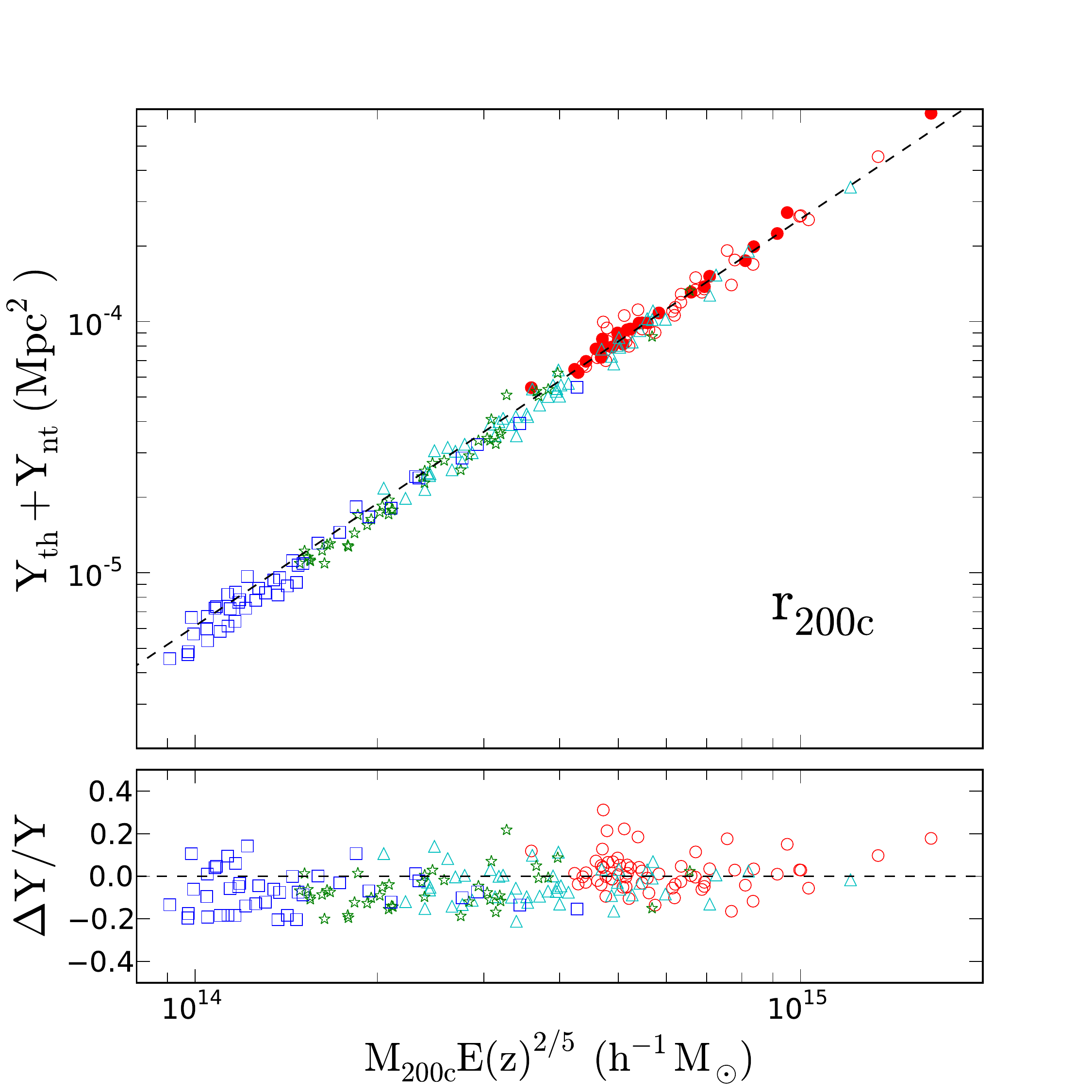}
  \caption{The $Y-M$ relations for both \yszth\ and \ysztot\ at redshifts $z=0, \ 0.5, \ 1, \ 1.5$ at $r_{500c}$ ({\it top}) and $r_{200c}$ ({\it bottom}). {\it Top panel of each plot:} the relation between total mass and the Comptonization parameter, Y. The dashed line shows the best-fit relation at $z=0$.  {\it Bottom panel of each plot:} the fractional deviation from the best-fit relation at $z=0$.}
\label{fig:YM} 
\end{figure*}

\section{Results}
\label{sec:results}

Using our simulations, we study the systematic biases in the SZE scaling relations due to the changing non-thermal energy and cluster dynamical state. We characterize the origin and evolution of deviations from the self-similar scaling relation at high redshift caused by incomplete thermalization of energy from major mergers.

\subsection{Effect of Major Mergers on the Y-M Scaling Relation}

We begin by characterizing the deviation of $Y$ from the best-fit relation for \ysztot\ at $z=0$ ($Y_{fit}$) as a function of time since the last major merger. In Table~\ref{tab:bestfit}, we show that $Y_{fit}$ agrees with the theoretically predicted self-similar mass scaling and redshift evolution and therefore herein we will use $Y_{fit}$ as a proxy for the self-similar relation.

Figure~\ref{fig:4in1} shows the fractional deviation of \yszth\ (red) at $r_{500c}$ and $r_{200c}$ from $Y_{fit}$, plotted against time since the last major merger, $t_{\rm merger}$, for the 65 $z=0$ simulated clusters and their most massive progenitors. Each cluster is traced from its latest $t_{\rm merger}$ value at $z=0$ back to just before the last major merger, and each line in the plot depicts the average \ysz\ values for all clusters as functions of $t_{\rm merger}$. Note that clusters with the same $t_{\rm merger}$ in the plot may not be at the same redshift. The dashed line depicts $Y_{fit}$ using the best fit normalization for \ysztot\ at $z=0$ (see Eq.~\ref{eq:ss}), which can be interpreted as a line of total energy. During the merger, 0 Gyr < $t_{\rm merger}$ < 2 Gyr, we see large deviations from the line of total energy, a trend also observed by \citet{wik08}. At this epoch the clusters are actively undergoing major mergers, producing large merger shocks (see N12 for more details). The shocks, which carry a large amount of pressure, result in an elevated \yszth. After 2 Gyr and 3 Gyr at $r_{500c}$ and $r_{200c}$, respectively, the shock has propagated outside of the region and the cluster is effectively a closed system. Over the next few Gyr the gas motions induced by the merger thermalize and we see a rise in the thermal $Y$ by $\sim 6\%$ for $r_{500c}$ and $\sim 8\%$ for $r_{200c}$. Observational works on the $Y-M$ relation to date \citep[e.g.,][]{sifon13, Saliwanchik2013} are still too small and noisy to reveal this trend, but our results agree with those from simulations by \citet{kay12}, \citet{battaglia11} and \citet{krause12}, who observed significant offset between regular and actively merging clusters in their gravitational effects-only simulation, where disturbed objects lie slightly below the $Y-M_{500}$ relation.  Specifically, \citet{battaglia11} quantified the dynamical states of their clusters with the ratio of kinetic to thermal energy, with a higher value corresponding to a more disturbed cluster. These results support our finding that there is a significant difference in the fraction of non-thermal energy between regular and disturbed clusters, and that this difference contributes to the apparent deviation from self-similar mass scaling.

As described in Section~\ref{subsec:e_of_nt}, we calculate the combined \ysztot\ to estimate the contribution from the missing non-thermal energy from the merger-induced gas motions. \ysztot\ is overplotted in blue on Figure~\ref{fig:4in1}.  For $Y$ measured within $r_{500c}$, \ysztot\ is able to recover the energy conservation in the ICM for epochs more than 1.5 Gyr after the merger. As mentioned above, the merger shock results in an excess of thermal energy that ultimately leaves the enclosed area, resulting in an overestimate of the total energy during the merger. For $Y$ measured within $r_{200c}$, there is an excess in \ysztot\ when the cluster is actively merging ($t_{\rm merger}$ < 1 Gyr) and then an underestimate in \ysztot\ for the next 2 Gyr. As the shock propagates outwards, it carries a significant amount of gas along with it to large radius, producing an apparent underestimate before the displaced ICM settles back within $r_{200c}$. This effect can also be seen between 1.75 and 2.25 Gyr within $r_{500c}$, but to a much lesser degree. Beyond 4 Gyr after the beginning of the merger, the system returns to a state of energy conservation and, by including the non-thermal energy contribution, we are able to recover the self-similar mass scaling and evolution to within $2\%$ for both radii.

\begin{figure*}[t]
\epsscale{1.1}
\plottwo{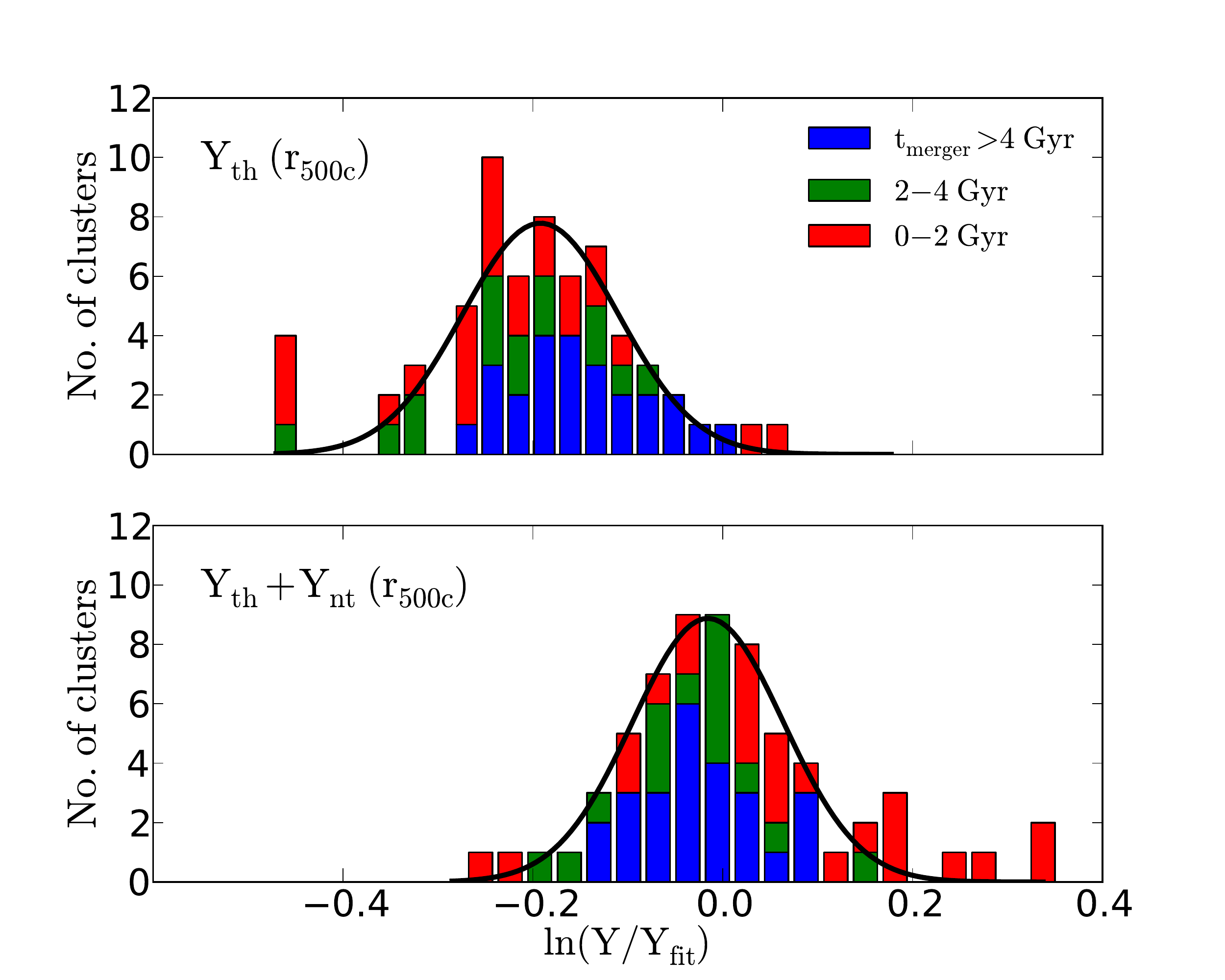}{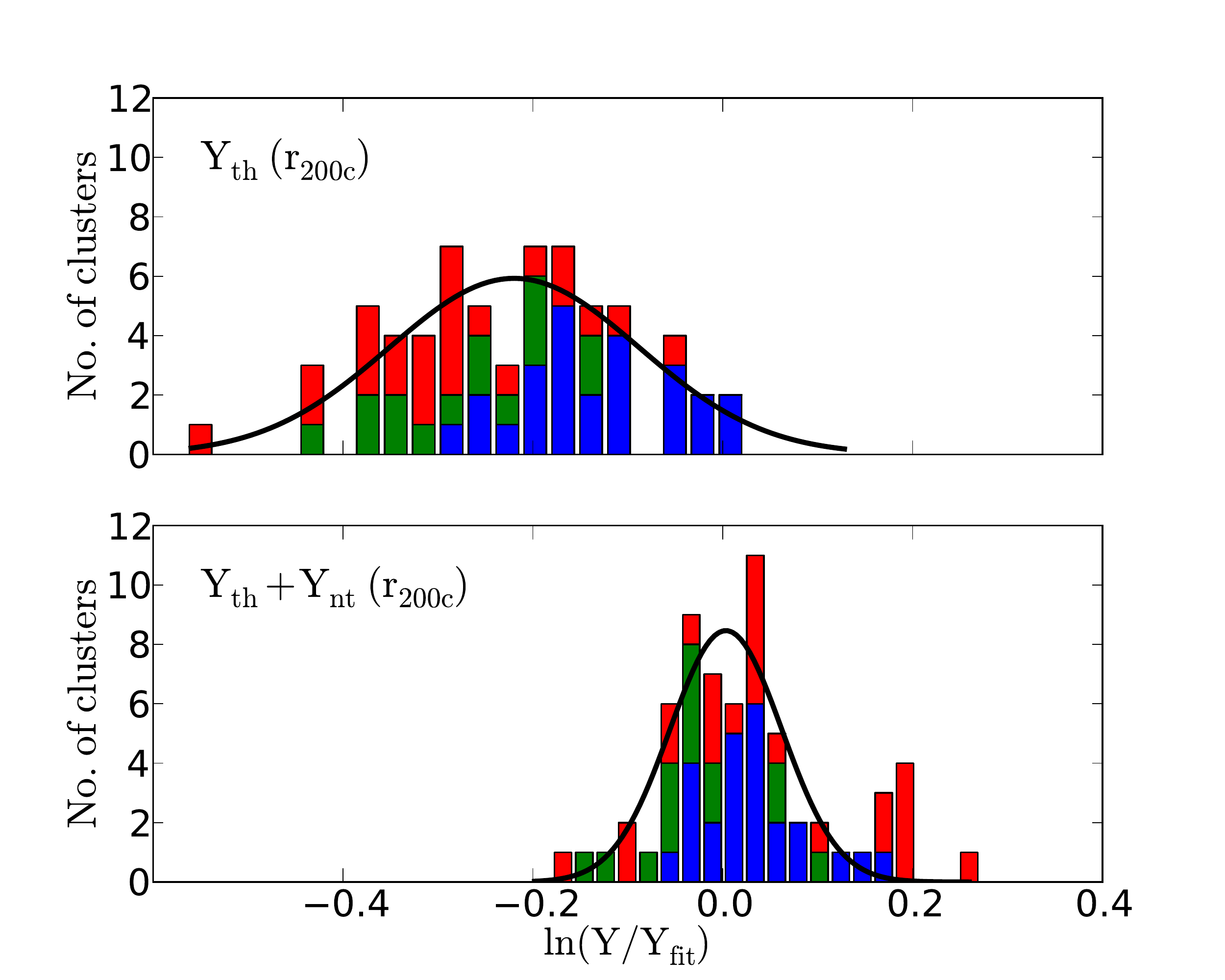}   
  \caption{Histograms of the fractional deviation of \yszth\ and \ysztot\ from $Y_{fit}$ at $z=0$ for clusters at $r_{500c}$ ({\it left}) and $r_{200c}$ ({\it right}), overplotted with the best-fit Gaussian curves. Each histogram is broken down into three $t_{\rm merger}$ bins: 0-2 Gyr ({\it red}), 2-4 Gyr ({\it green}) and > 4 Gyr ({\it blue}).}
\label{fig:hist} 
\end{figure*}

\subsection{Effect on the $Y-M$ Scaling Relation}
\label{subsec:overview}

The above dynamical state dependence and evolution has significant implications for the observable-mass, $Y-M$, relation.  Figure~\ref{fig:YM} shows the $Y-M$ scaling relations, both with and without $Y_{nt}$, of the simulated clusters at $r_{500c}$ (left panels) and $r_{200c}$ (right panels), with the sample spanning redshifts 0, 0.5, 1 and 1.5 (shown as circles, triangles, stars and squares, respectively). We fit the scaling relation  (Eq.~\ref{eq:ymod}) to the samples of simulated clusters at each redshift. The dashed line depicts $Y_{fit}$ as fit for the $z = 0$ sample for each respective $Y$ and radius. 

Table~\ref{tab:bestfit} also reports the best-fit slopes and logarithms of the best-fit normalizations for scaling relations at each redshift. There is no significant redshift evolution in the normalization of \yszth\ within $r_{500c}$, in agreement with \citet{nagai06}, while the normalization of \yszth\ within $r_{200c}$ shows a deviation from the self-similar redshift evolution model towards higher redshift. This departure from self-similarity can be clearly seen in the green ($z=1.0$) and blue ($z=1.5$) points, which lie below the best-fit line for $z=0$. The combination of (a) cluster outskirts having larger fractional non-thermal pressure \citep[e.g.][]{nelson14b,shi14a,shi14b} and (b) the expectation that clusters at higher redshift are growing more rapidly leads to a large fractional non-thermal pressure for a given measured $Y$ at high $z$, and in turn results in an overestimate in mass compared to the self-similar prediction. However, when we include the non-thermal energy contribution from bulk motions, the normalization of \ysztot\ is more self-similar at all redshifts and radii. 

We also calculate the scatter in $Y$ at a given mass at each epoch as 
\begin{equation}
\sigma = \sqrt{\frac{\sum \limits^N (\ln Y - \ln Y_{fit})^2} {N-1}}
\end{equation}
where $N$ is the size of the sample, $Y$ is either \yszth\ or \ysztot\ as measured directly from the simulation, and  $Y_{fit}$ is the $Y$ value predicted by the best-fit relation for each respective $Y$ and radius for a given cluster's mass and redshift, with redshift evolution held at the self-similar value. Table~\ref{tab:bestfit} reports the scatter for both thermal and combined $Y$. The quoted errors represent the $68 \%$ confidence interval, estimated using the bootstrap method with 10,000 random samples.  The scatter in \yszth\ does not evolve with redshift at $r_{500c}$, consistent with the findings of \citet{Sembolini2012} based on their non-radiative simulation. We do, however, see a trend towards larger scatter in the outer region of clusters, where there is a high non-thermal energy fraction due to longer relaxation time in the low-density region.  At most epochs, the scatter in \ysztot\ , on the other hand, is systematically smaller by $\sim 20-30\%$ than that of \yszth\ alone and comparable at $r_{500c}$ and $r_{200c}$.

In Figure~\ref{fig:YM}, we also examine the effects of dynamical state on the scaling reaction.  Our relaxed subsample (defined using clusters at $z=0$ that have not experienced any major mergers in the previous 4 Gyr) are shown in solid circles. The scatter for this subsample is also given in the last row of Table~\ref{tab:bestfit}. The relaxed subsample always has significantly lower scatter than the entire sample at the same radius and redshift, for both \yszth\ and \ysztot. This result agrees with \citet{battaglia11}, who observed that the least disturbed clusters, defined as those with the smallest ratios of kinetic to thermal energy, show the lowest scatter ($11\%$ in their AGN simulation). If we compare the scatter of the thermal \yszth\ for the entire sample to that of the combined \ysztot\ for the subsample, the reduction in scatter is as large as $\sim 50 \%$ for both radii. In Section~\ref{subsec:massacc}, we compare the effects of major mergers to more general accretion in driving the scatter in the relation.

\subsection{Distribution of Scatter in the $Y-M$ relation}

We examine the relative contributions to the scatter in the observable-mass relation from clusters in different stages of major mergers.  In Figure~{\ref{fig:hist}, we show stacked histograms of the logarithmic fractional deviations of \ysz\ from the self-similar predictions for different $t_{\rm merger}$ bins, with $Y_{th}$ in the top panels and $Y_{th} + Y_{nt}$ in the bottom panels. The $Y$ values are normalized to $Y_{fit}$ calculated as \ysztot\ for the $z = 0$ clusters at each respective radius. The cluster inner regions ($r_{500c}$) and outskirts ($r_{200c}$) are shown on the left and right, respectively. The actively merging systems are shown in red, having experienced a major merger in the last 2 Gyr. The most relaxed clusters, with $t_{\rm merger}$ > 4 Gyr, are shown in blue and the intermediate stage clusters shown in green. For reference, we have overplotted a Gaussian fit, shown as black curves. 

For $Y_{th}$ alone, we find that the largest deviations come from actively merging clusters ($0-2$~Gyr), particularly towards lower values of $Y$. Consistent with Figure~\ref{fig:4in1}, we find that the relaxed subsample tends to lie at the higher end of the scatter distribution at both radii due to the smaller non-thermal energy fraction of the ICM and therefore larger $Y_{th}$ value for a given mass. Likewise, the intermediate dynamical state is between the relaxed and actively merging states. The radial dependence of the scatter distribution shown in Table~\ref{tab:bestfit} is also clear in top panels, with a wider distribution in the cluster outskirts where the fractional missing non-thermal energy contribution is larger.

When we add $Y_{nt}$, we see a tightening of the scatter and a weakening of the dynamical state dependence at both radii. However, the size of the scatter still shows strong dynamical state dependence, with the relaxed subsample exhibiting the smallest scatter and the actively merging systems the largest. There is also a tail in the actively merging distribution towards high $Y$. This is the overestimation shown in Figure~\ref{fig:4in1} due to the excess energy from the merger shock during that time period.

\begin{figure*}[t]
\epsscale{0.90}
\plottwo{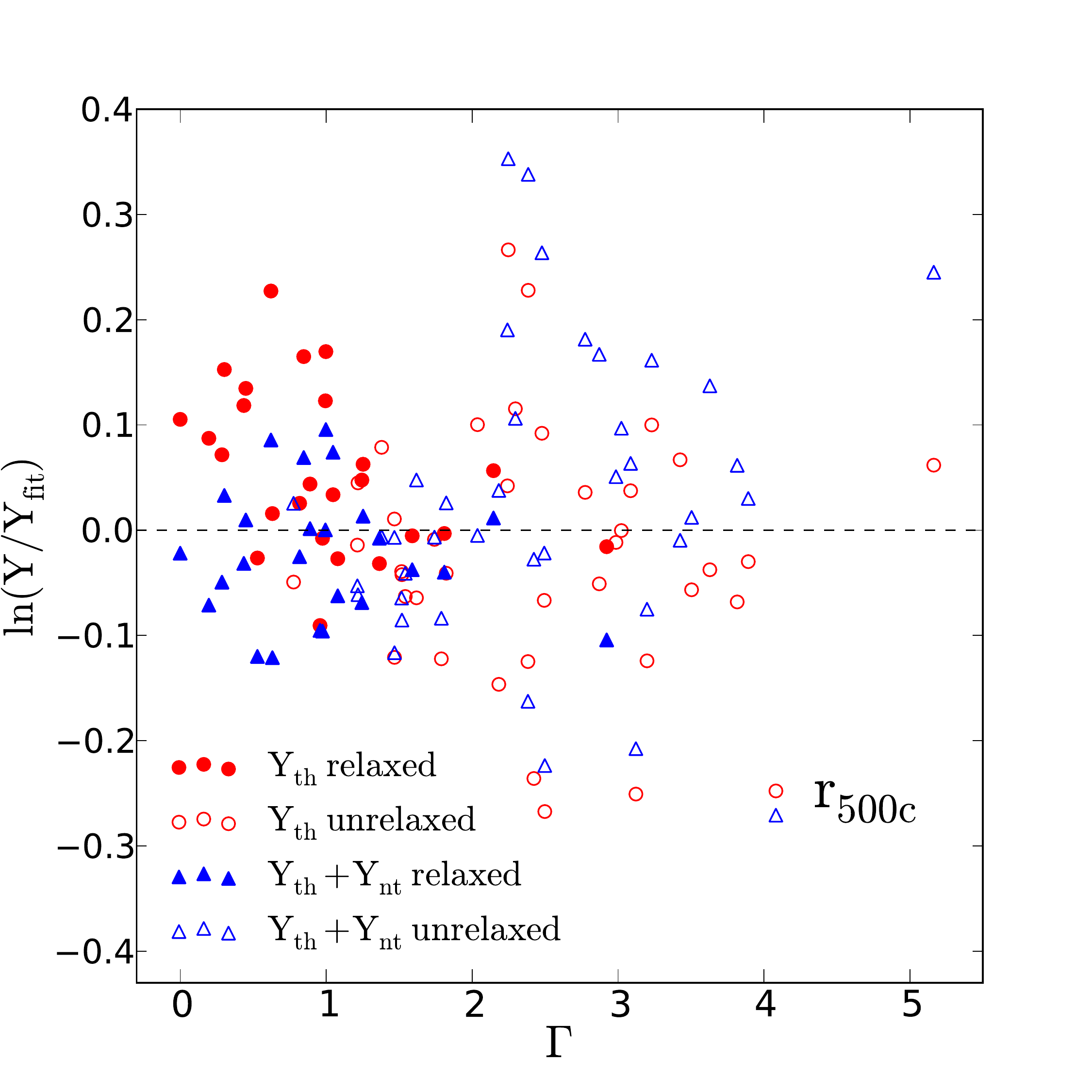}{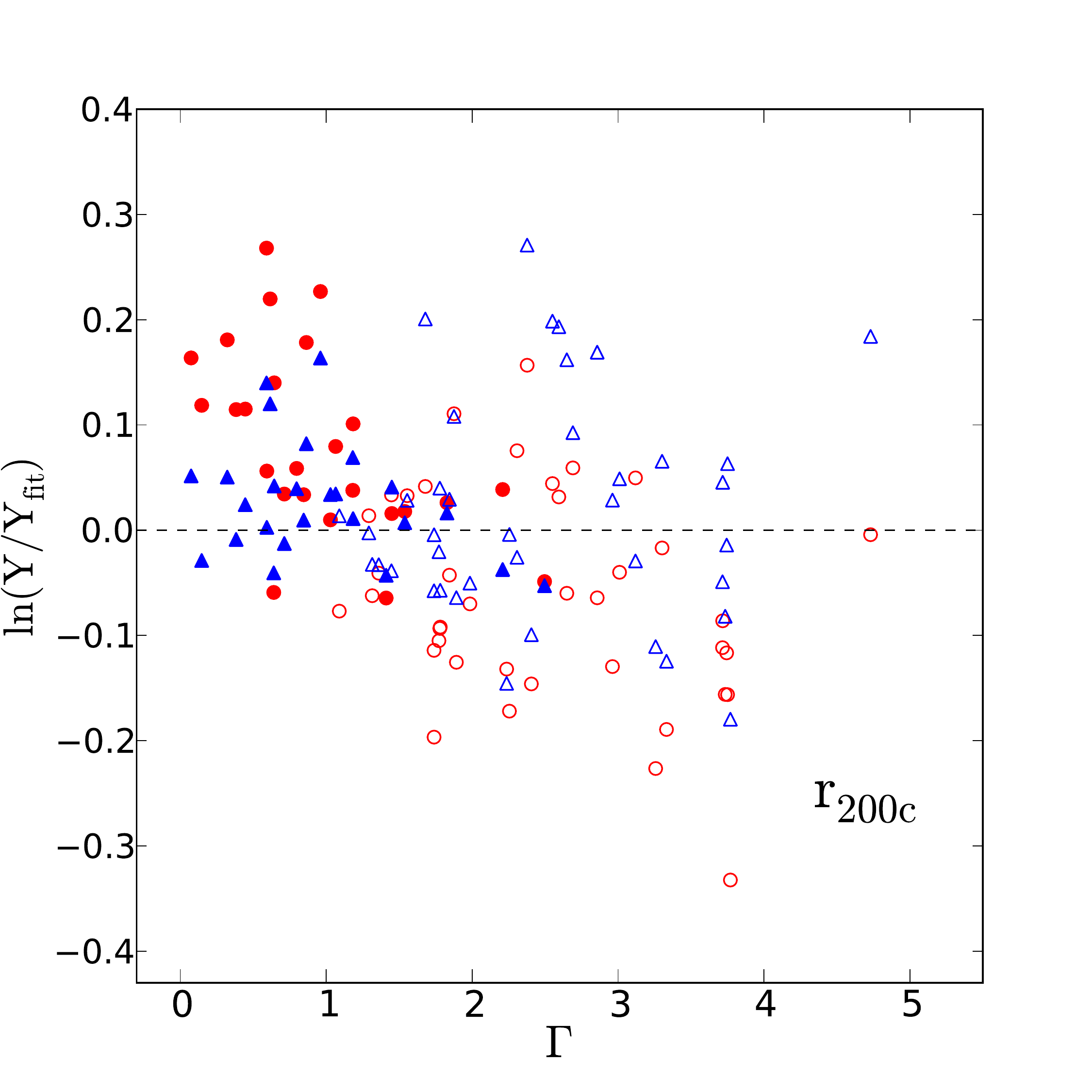}   
  \caption{Logarithmic deviation of \yszth\ and \ysztot\ at $z=0$ from $Y_{fit}$, calculated separately for \yszth\ and \ysztot, for clusters at $r_{500c}$ ({\it left}) and $r_{200c}$ ({\it right}), plotted as functions of $\Gamma$. The blue triangles and red circles show \yszth\ and \ysztot\ respectively. Filled symbols are the clusters with $t_{\rm merger}$ > 4 Gyr, and open symbols represent the remaining clusters.}
\label{fig:mass_acc} 
\end{figure*}

\subsection{Effect of Other Accretion Processes}
\label{subsec:massacc}

Since clusters do not accrete mass only through major mergers, we investigate whether the observed dynamical state dependence in the $Y-M$ scaling relation is due primarily to major mergers or the general accretion rate, regardless of accretion mode (including minor mergers and/or smooth accretion). In this section, we generalize our analysis to total mass accretion, and eliminate clusters that undergo recent major mergers. The remaining clusters allow us to discern the effect of non-major merger accretion in that period.

To quantify the total mass accretion of each cluster, we define the quantity $\Gamma$ following \citet{diemer14}:
\begin{equation}
\Gamma \equiv \Delta \log M/\Delta \log a
\end{equation}
where $a$ is the expansion factor. Here, $\Gamma$ is defined using the masses of the main progenitor at $z=0.5$ and its descendant at $z=0$ within $r_{500c}$ or $r_{200c}$. In Figure~\ref{fig:mass_acc}, we show $\ln$ ($Y$/$Y_{fit}$) at $z=0$ for each cluster plotted as a function of $\Gamma$ for both $r_{500c}$ and $r_{200c}$. In this figure, $Y_{fit}$ is calculated separately for \yszth\ and \ysztot\ at $z = 0$ to better highlight the relative distribution of scatter. \yszth\ is shown as red circles, and \ysztot\ as blue triangles. The clusters at $z=0$ without major mergers in the last 4 Gyr are highlighted as filled shapes. 

There is a clear trend of \yszth\ with $\Gamma$, with more slowly accreting systems having higher $Y_{th}/Y_{fit}$ and more rapidly accreting ones having lower $Y_{th}/Y_{fit}$. The lower $\Gamma$ clusters are accreting mass more slowly than the higher $\Gamma$ counterparts and, therefore, contain a more thermalized ICM, resulting in a higher \yszth\ for a given mass. This trend is stronger in the outskirts, within $r_{200c}$, where the non-thermal energy fraction in unrelaxed systems is larger and the clusters at high $\Gamma$ have lower values of \yszth.  However, when the non-thermal energy contribution is included, \ysztot\ exhibits little to no dependence on $\Gamma$ at all radii. The vast majority of clusters with large $\Gamma$ are not highlighted, indicating that clusters that accrete significant amounts of mass do so mostly through major mergers. 

There are several clusters with $\Gamma > 1.5$ that are accreting rapidly (representing clusters that have undergone strong mass accretion), but no major merger (depicted by the filled shapes). However, these clusters lie very close to the self-similar relation and contribute little to the scatter at large $\Gamma$ for either \yszth\ or \ysztot. The high $\Gamma$ clusters with the larger deviations from self-similarity are the un-highlighted, recently merged systems. Thus, we conclude that the dominant dynamical state effects on the scatter in the $Y-M$ relations are due to major mergers and other forms of accretion are subdominant. However, it is likely that in the most relaxed systems, non-thermal pressure injection from ongoing smooth accretion and/or minor mergers are responsible for residual bias and scatter in the relation. This can be seen in Table~\ref{tab:bestfit} as the reduction of scatter for the relaxed subsample with the addition of the non-thermal energy contribution, especially at large radii.

\section{Discussion and Conclusions}
\label{sec:conc}

In this work, we have studied the relation between cluster dynamical state and scatter and evolution in $Y-M$ scaling relation, using a mass-limited sample of massive galaxy clusters from the {\it Omega500} high-resolution hydrodynamic simulation. To characterize the effects of mergers and dynamical states of clusters, we analyzed non-radiative simulations in which the ICM is heated by gravitational heating processes alone and used merger trees to follow the evolution of individual clusters during and after mergers and trace their mass accretion histories. Our main results are summarized as follows:

\begin{enumerate}
\item Scatter in the $Y-M$ scaling relation arises largely from the evolution of missing non-thermal pressure sourced by gas motions in galaxy clusters. The thermal SZE signal increases systematically after a major merger, as the energy from random gas motions decays into thermal energy. 
\item We add the non-thermal energy contribution provided by random gas motions and find that the scaling relation for the combined $Y$ exhibits little to no systematic evolution and its scatter is reduced by $20-30\%$ at each redshift we examined.
\item The intrinsic log-normal scatter in the $Y-M$ relation of a relaxed subsample at $z=0$ is about a third smaller than that of the entire sample, with a logarithmic 1-$\sigma$ scatter of only 7.0$\%$ and 8.5$\%$ compared to 11.4$\%$ and 12.3$\%$ in \yszth\ for the full $z=0$ sample at $r_{500c}$ and $r_{200c}$ respectively. 
\item We also examine the effect of other mass accretion processes on scatter around the scaling relation, but conclude that their contribution is subdominant compared to that of major mergers. Thus, we conclude that the main effects of dynamical state on scatter are driven by major mergers.
\end{enumerate}

Our results based on cosmological simulations are in general agreement with the results of the idealized merger simulations by \citet{wik08} regarding the impact of mergers on the SZE signal. However, there are also important quantitative differences that highlight the advantages of cosmological simulations. For example, \citet{wik08} reported a steep rise (almost by a factor of 2) in $Y$ due to mergers.  While we also see a similar rise in Figure~\ref{fig:4in1}, the increase in $Y$ due to merger is more modest with $\lesssim$ 80\% between the post- and pre-merger values. In idealized simulations, a merger exists in isolation. In cosmological simulations, a merger occurs alongside ongoing smooth accretion and more ubiquitous minor mergers, which contribute non-negligible non-thermal energy to the ICM and result in a lower value of $Y$.

The post-merger evolution in $Y$ presented herein closely echoes the findings in N12 for the hydrostatic mass bias. They also find large deviations in the hydrostatic mass estimate due to merger shocks for the actively merging systems ($t_{\rm merger}$ < 2 Gyr) and systematic bias in the mass estimate due to missing non-thermal pressure for relaxed systems ($t_{\rm merger}$ > 4 Gyr). Similarly, by accounting for the non-thermal pressure, there is a significant improvement in agreement between the hydrostatic mass estimate and the true mass. It is interesting, although unsurprising, to find such close correlation in the timescales related to both the hydrostatic mass bias and the scatter in the $Y-M$ scaling relation, highlighting the impact and importance of mergers and the unobserved non-thermal energy contribution in galaxy clusters. It is worth highlighting that this co-evolution has implications for observational studies of the $Y-M$ using the hydrostatic mass estimate. The scatter, particularly for unrelaxed clusters, may be affected along both axes by the presence of non-thermal pressure contributions. 

There are additional sources of non-thermal pressure support not included or addressed in our simulation. While the inclusion of cooling and star formation does not affect the size of the scatter in the $Y-M$ relation \citep{nagai06}, strong energy feedback by AGN may enhance the scatter by a few percent, especially at high redshift where star formation and AGN are active \citep{battaglia11, fabjan11}.  Additionally, magnetic fields and cosmic rays can provide additional non-thermal pressure \citep[e.g.,][]{lagana10}, but their contributions are expected to be negligible: the ratio of cosmic ray pressure to total pressure is constrained to $\lesssim 1\%$, set by the $\gamma$-ray observations of {\em Fermi}-LAT \citep{Fermi2013}. Similarly, the typical magnetic field strength of $\lesssim 10 \mu\mathrm{G}$ in the ICM corresponds to magnetic pressure fraction of $\lesssim 1\%$.  However, plasma effects can amplify gas turbulence and provide extra non-thermal pressure support \citep{parrish12}, while physical viscosity, on the other hand, can lower the non-thermal pressure support by decreasing the level of gas turbulence. Moreover, the electron-ion equilibration timescale is expected to be longer than to the age of the universe in the low density galaxy cluster outskirts ($r\gtrsim r_{200c}$), which causes the electron temperature and hence the thermal SZE signal to be lower than the prediction of hydrodynamical simulation which assumes the local thermal equilibrium of electrons and ions. However, the effect of the non-equilibrium electrons on the integrated thermal SZE signal (\yszth) is of order $\lesssim 3\%$ and $\lesssim 5\%$ at $r_{500c}$ and $r_{200c}$, respectively \citep{rudd09,avestruz14}, which is much smaller than the effect of merger-induced non-thermal pressure discussed in this work. The results presented in this paper based on hydrodynamical simulations serve as a baseline for further studies of these effects.

Recent observational studies of the $Y-M$ scaling relation using various observational mass measurements, such as hydrostatic mass estimate and gravitational lensing, have found an intrinsic log-normal scatter of 20-30\% \citep[e.g.][]{Marrone2009, Marrone2012, Saliwanchik2013}. This is twice the size of the intrinsic scatter measured from simulations, potentially signifying that additional sources of scatter are not accounted for in the simulations. While accounting for the non-thermal pressure associated with turbulent motions in the ICM significantly reduces the scatter, and evolution in the $Y-M$ relation, especially in relaxed clusters, gas acceleration can have a non-negligible effect, especially in unrelaxed systems \citep{suto13, lau13, nelson14a}.  Focusing on a sub-sample of relaxed clusters (which can be defined, for example, using morphological measure, such as centroid shift, ellipticity, or power ratios, as a proxy for the cluster's dynamical state) can help minimize the effect of this irreducible bias associated with gas acceleration. In practice, there is also additional scatter due to projection effects, which could be significant especially for low-mass clusters and groups \citep{white02, hallman07}.  We will investigate these effects in our upcoming work. In observations, one also does not know a priori the true values for $M_{500c}$ or $r_{200c}$ as we do in simulations. \citet{Kravtsov2006} present an iterative algorithm for simultaneously fitting for $r_{500c}$ and $M_{500c}$ using scaling relations. Estimating the appropriate radii for the use of $Y_{th}$ and $Y_{th} + Y_{nt}$ in future observations could also be achieved using this technique. Note, however, that any biases in the calibration of the $Y-M$ relation can introduce uncertainties in $M_{500c}$ and hence $r_{500c}$.

Our results have an important implication for cosmological constraints derived from SZE cluster surveys. Since \ysztot\ is a better proxy for mass than \yszth, our work opens up a possibility to further improve the current robust mass proxy \yszth\ by measuring and correcting for the missing non-thermal pressure component \ysznt. For example, the upcoming {\it ASTRO-H} X-ray observation satellite will provide the first direct measurement of internal gas motions in clusters, which in turn should provide observational constraints on the non-thermal energy content in clusters.  However,  {\it ASTRO-H} measurements will be limited to the inner regions ($r\lesssim r_{2500c}=0.5r_{500c}$) of nearby clusters, unless one can dedicate a very long exposure of order $\gtrsim 1$~Msec \citep{nagai13}. Alternatively, the next generation of high-resolution, multi-wavelength SZE imaging observations may enable us to measure the non-thermal energy component through the kinematic SZ effect signal generated by internal gas motions in clusters \citep{sunyaev03,nagai03}. Recent theoretical works have found strong yet simple relations between the non-thermal pressure and the more easily observable thermal pressure profiles, potentially supplying an alternate route for estimating the energy contribution from random motions in galaxy clusters \citep{nelson14b,shi14a,shi14b}.

\acknowledgments 
We thank Stefano Andreon, August Evrard and the anonymous referee for comments on the manuscript and Erwin Lau for discussions during the early stage of this work. This work was supported in part by NSF grants AST-1412768 \& 1009811, NASA ATP grant NNX11AE07G, NASA Chandra grants GO213004B and TM4-15007X, the Research Corporation, and by the facilities and staff of the Yale University Faculty of Arts and Sciences High Performance Computing Center.  LY also acknowledges support from the Yale Science, Technology and Research Scholars (STARS) II Program.

\bibliography{ms}

\end{document}